\documentclass[10pt,journal,compsoc]{IEEEtran}

\ifCLASSOPTIONcompsoc
  \usepackage[nocompress]{cite}
\else
  \usepackage{cite}
\fi

\usepackage{graphicx}
\usepackage{comment}
\usepackage{amsmath,amssymb} % define this before the line numbering.

\usepackage{multirow}
\usepackage{multicol}
\usepackage{floatrow}
\usepackage{placeins}
\usepackage[linesnumbered,ruled,vlined]{algorithm2e}
\SetAlgoSkip{} % Remove extra spacing
\SetInd{0.5em}{1em} % Adjust indentation if needed
\usepackage{bbding}

\usepackage{soul,tikz,color,xcolor,hyperref}% Make Orcid icon
\usepackage{pifont}
\usepackage{subfigure}
\usepackage{booktabs}
\usepackage{makecell}
\usepackage{xspace}

\usepackage{listings}
\usepackage{tcolorbox}
\newcommand{\printRealNewlines}[1]{%
  \texttt{\lstset{style=jsonStyle,basicstyle=\ttfamily\small}%
    \begingroup
      \ttfamily\small
      \expandafter\replaceNewlines#1\relax
    \endgroup}%
}
\def\replaceNewlines#1\n#2\relax{%
  #1%
  \ifx\relax#2\relax
  \else
    \\        % 这里产生真正的换行
    \replaceNewlines#2\relax
  \fi
}
\lstdefinestyle{jsonStyle}{
    basicstyle=\small\ttfamily,
    breaklines=true,
    breakatwhitespace=true,
    columns=flexible,
    keepspaces=true,
    showstringspaces=false,
    frame=single,
    backgroundcolor=\color{gray!10},
}

\lstdefinestyle{jsonStyle2}{
    basicstyle=\small,
    breaklines=true,
    breakatwhitespace=true,
    columns=flexible,
    keepspaces=true,
    showstringspaces=false,
    frame=single,
    backgroundcolor=\color{gray!10},
}

\definecolor{lime}{HTML}{A6CE39}
\DeclareRobustCommand{\orcidicon}{%
    \begin{tikzpicture}
    \draw[lime, fill=lime] (0,0) 
    circle [radius=0.16] 
    node[white] {{\fontfamily{qag}\selectfont \tiny ID}};    \draw[white, fill=white] (-0.0625,0.095) 
    circle [radius=0.007];    \end{tikzpicture}
    \hspace{-2mm}}
\foreach \x in {A, ..., Z}{%
    \expandafter\xdef\csname orcid\x\endcsname{\noexpand\href{https://orcid.org/\csname orcidauthor\x\endcsname}{\noexpand\orcidicon}}
    }

\begin{document}
\title{BioinfoMCP: A Unified Platform Enabling MCP Interfaces in Agentic Bioinformatics}

% \author{Michael~Shell,~\IEEEmembership{Member,~IEEE,}
%         John~Doe,~\IEEEmembership{Fellow,~OSA,}
%         and~Jane~Doe,~\IEEEmembership{Life~Fellow,~IEEE}% <-this % stops a space
% \author{Chong~Mou, Jian~Zhang
\author{Florensia~Widjaja$^{1,\dagger}$, Zhangtianyi Chen$^{1,\dagger}$, Juexiao Zhou$^{1,*}$
\thanks{
$^1$School of Data Science, The Chinese University of Hong Kong, Shenzhen (CUHK Shenzhen), Shenzhen 518172, China\\
$^\dagger$These authors contributed equally.\\
$^*$Corresponding author. e-mail: juexiao.zhou@gmail.com\\
}}


\markboth{}%
\\
\IEEEtitleabstractindextext{%
\begin{abstract}
Bioinformatics tools are essential for complex computational biology tasks, yet their integration with emerging AI‑agent frameworks is hindered by incompatible interfaces, heterogeneous input–output formats, and inconsistent parameter conventions. The Model Context Protocol (MCP) provides a standardized framework for tool–AI communication, but manually converting hundreds of existing and rapidly growing specialized bioinformatics tools into MCP‑compliant servers is labor‑intensive and unsustainable. Here, we present BioinfoMCP, a unified platform comprising two components: BioinfoMCP Converter, which automatically generates robust MCP servers from tool documentation using large language models, and BioinfoMCP Benchmark, which systematically validates the reliability and versatility of converted tools across diverse computational tasks. We present a platform of 38 MCP‑converted bioinformatics tools, extensively validated to show that 94.7\% successfully executed complex workflows across three widely used AI‑agent platforms. By removing technical barriers to AI automation, BioinfoMCP enables natural‑language interaction with sophisticated bioinformatics analyses without requiring extensive programming expertise, offering a scalable path to intelligent, interoperable computational biology.
\end{abstract}

}

\maketitle

\IEEEdisplaynontitleabstractindextext

% \fi
\IEEEpeerreviewmaketitle

\section{Introduction}
The bioinformatics landscape is characterized by an extensive ecosystem of specialized tools designed for diverse analytical tasks that serve critical functions in genomics \cite{noauthor_genome_nodate}, proteomics \cite{wetie_protein-protein_2014, protein_interaction_network}, and molecular biology \cite{genomic_molecule_bio}, and so on. Each tool typically operates as a standalone application with unique input-output formats, command-line interfaces, and computational requirements, and is also designed for a specialized purpose \cite{genomic_file_format}. These tools were then utilized in a strategic sequence of computational steps to produce interpretable results in domains of genomic analysis \cite{human_genome_sequence}, structural bioinformatics \cite{structural_genomics}, and also in computational methods such as data and text mining
 \cite{zhao_text_mining_2021}, phylogenetics \cite{wu_phylogenomic_2012}, or in population studies \cite{rivas_efficient_2025}. Prevalent end-to-end tasks, which oftentimes are called pipelines, are then executed encompassing various datasets, such as whole genome sequencing (WGS) \cite{human_genome_sequence}, Chromatin Immunoprecipitation Sequencing (ChIP-seq) \cite{chipseq_review, Chipseq_macs}, RNA sequencing (RNA-seq) \cite{rnaseq_cell_differentiation, rnaseq-survey}, single-cell RNA-seq (scRNA-Seq) \cite{scRNA_pipeline, scRNA_data_tech}, and also other widely-utilized sequencing studies. The accomplishments of these sequencing studies bring out the actionable biological insights, predictive biomarkers, therapeutic targets, and personalized treatment strategies that have contributed to significant progress in precision medicine \cite{LLM_medicine} and drug discovery \cite{zhang_role_drug_dis}.

The field of artificial intelligence (AI) has experienced significant advancement in recent years, most notably through the emergence of large language models (LLMs) such as OpenAI's ChatGPT \cite{gpt4o} and Anthropic's Claude \cite{claude}. These developments have profoundly altered human-machine interaction paradigms, with ongoing research and development suggesting sustained momentum in this domain. Even in August 2025, OpenAI also released the latest GPT-5, which was acknowledged to have the knowledge capacity of a postgraduate student \cite{gpt5}. Rapid breakthroughs in AI express a need for many fields to harness its power. However, domain-specific tools from bioinformatics, though highly valuable, have struggled to be integrated into these cutting-edge models. For instance, established bioinformatics tools were primarily designed for direct human interaction rather than programmatic access, resulting in incompatible data formats, limited API availability, and workflow structures that impede their integration into AI-driven analytical pipelines \cite{genomic_file_format, genomic_reproducibility}.

Due to the steep learning curve and tedious procedures of using these lack-of-standardization software, researchers have tried to utilize AI agents to make these bioinformatic pipeline analyses more autonomous and time and energy-efficient, such as AutoGPT \cite{autogpt}. However, its general-purpose coverage ability significantly decreases the robustness of their responses and makes redevelopment a challenging task \cite{autoba}.  In order to address the limitations that general AI agents faced, some specialized AI agents that focused specifically on addressing the needs of bioinformaticians were released. These systems encompass a spectrum from semi-automated web-based platforms such as iDEP \cite{idep} and ICARUS \cite{10.1093/bioinformatics/icarus}, to fully autonomous frameworks including BioPlanner \cite{bioplanner}, BioAgents \cite{bioagents}, MCPMed \cite{mcpmed}, AutoBA \cite{autoba}, BRAD \cite{brad}, and Spatial Agent \cite{spatialAgent}, which provide specialized computational capabilities for researchers undertaking sophisticated bioinformatics workflows.

Despite advances in AI agents, scientists must still determine how best to integrate these systems into their research workflows, particularly in linking the diverse bioinformatics tools required for pipeline analyses, each with distinct procedures and requirements \cite{LLM_medicine, rnaseq-survey}. Meanwhile, the majority of existing bioinformatics tools lack standardized interfaces, and the rapid emergence of new specialized tools further compounds this challenge. As these tools often adopt distinct architectures, file formats, and operational conventions, AI agents face significant obstacles in directly invoking them without extensive, tool‑specific adaptation \cite{genomic_file_format, bio_tool_rec_system}. This fragmentation not only limits the scalability of AI‑driven workflows but also slows the adoption of automation in computational biology, as each integration effort requires substantial manual engineering. Consequently, there is a critical need for a general mechanism that can bridge AI agents with this continually expanding and heterogeneous landscape of bioinformatics software.

Model Context Protocol (MCP) is a general-purpose protocol that acts as a standard for agentic-tool communication \cite{anthropic2024model}. MCP provides a unified connection format between tools or applications and an extensive set of AI Agents, which are also called MCP Hosts. These MCP Clients, which have attached an MCP Client to them like Claude Desktop \cite{claude} or Cursor \cite{cursor}, can seamlessly connect to external tools that were packaged into MCP servers using a standardized protocol, hence making it easy for AI agents to run commands on these tools.

Converting bioinformatics tools into MCP servers addresses these fundamental integration challenges by providing a standardized communication layer to any AI agent that has an MCP host integrated into it. MCP servers enable tools to expose their functionality through consistent interfaces, allowing AI assistants and automated systems to seamlessly interact with diverse bioinformatics applications. This standardization facilitates the creation of intelligent workflows where tools can be dynamically selected and chained based on analytical requirements rather than technical compatibility constraints. Furthermore, MCP integration enables natural language interaction with complex bioinformatics tools, making advanced analyses accessible to researchers without deep technical expertise while maintaining the specialized capabilities that make these tools valuable to the scientific community.

However, manual conversion of bioinformatics tools into MCP servers would be impractical given the scale and diversity of the ecosystem. With hundreds of specialized tools across genomics, proteomics, and molecular biology, each requiring a deep understanding of their specific interfaces, parameters, and data formats, manual conversion would be time-consuming and error-prone. Moreover, the bioinformatics field continuously evolves with new tools and updated versions of existing ones, making manual approaches unsustainable \cite{huang_bioinformatics_2009, noauthor_genome_nodate}. With an automatic conversion system, it can systematically analyze tool documentation, command-line interfaces, and input/output specifications to generate standardized MCP server implementations at scale. This automation ensures consistency across conversions, reduces development time from months to minutes per tool, and enables rapid adaptation to tool updates and new releases, making MCP integration feasible for the entire bioinformatics ecosystem and demolishing the steep learning curve for scientists in utilizing these useful-but-standalone bioinformatician tools. 

Here, we present the BioinfoMCP platform, which tackles three fundamental barriers limiting bioinformaticians' productivity: \textbf{1) Fragmentation and Incompatibility of Bioinformatics Tools} - hundreds of specialized standalone tools with incompatible interfaces, diverse input/output formats, and inconsistent parameter naming conventions create substantial integration barriers that require extensive manual effort to overcome; \textbf{2) Lack of AI Agent Integration} - existing bioinformatics tools were designed for direct human interaction rather than programmatic access by AI agents, lacking the standardized APIs and communication protocols necessary for seamless integration with modern AI-driven workflows; and \textbf{3) Manual Conversion Bottleneck} - manual conversion of each bioinformatics tool into an MCP server would require substantial development time per tool, making this approach unsustainable and impractical given the vast, continuously evolving ecosystem of bioinformatics software.

\begin{figure*}[!htb]
    \centering
    \includegraphics[width=1\columnwidth, alt={The top part (a) shows the flow of BioinfoMCP Converter, from preparation of documentation, backbone initialization, execution of code generation, extraction, and evaluation, and finally containerize the tools. The bottom part (b) shows how the tools got tested with AI Agents, and on complex tasks which then return the ran tools, output files, and a comprehensive summary.}]{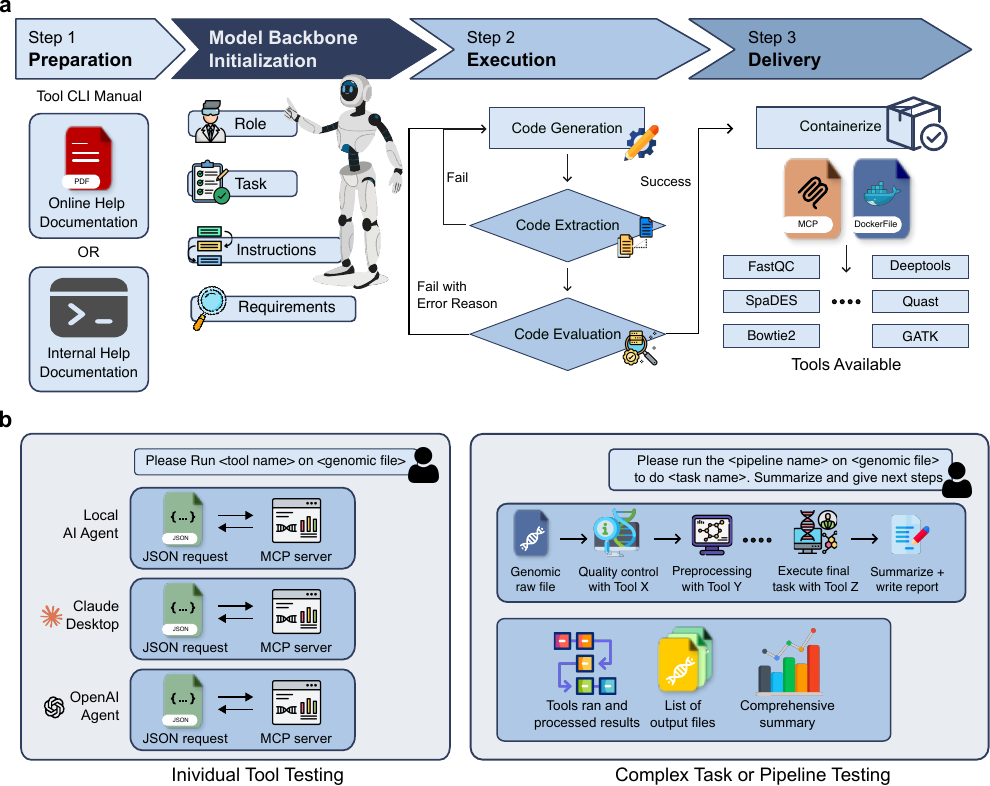}
    \caption{\emph{Design of BioinfoMCP}, which consists of two parts: a) BioinfoMCP Converter and b) BioinfoMCP Benchmark.}
    \label{fig:workflow}
\end{figure*}

Considering the issues that were observed, as can be seen in Figure \ref{fig:workflow}, the BioinfoMCP was created with two main branches for development, which are the \textbf{BioinfoMCP Converter}, that works as a script to automatically convert bioinformatics tools into a robust executable MCP server -- described further in Section \ref{sec2-converter}, and also the supporting \textbf{BioinfoMCP Benchmar}k, that manually curated set of test cases for MCP servers to analyze the robustness and versatility of BioinfoMCP converter-converted tools across different tasks -- described further in Section \ref{sec2-benchmark}. 

\section{Methods}

\subsection{Design of BioinfoMCP Converter}
\label{sec2-converter}

\begin{algorithm*}[t]
  \label{algo:converter}
  \caption{BioinfoMCP Converter}
  \KwIn{tool name $T$, help manual path $M$, help manual flag $h\in\{0,1\}$, output dir $O$, \\
        LLM model $\mathcal{M}$, API key $\mathcal{K}$}
  \KwOut{Dockerised MCP server ready for an AI Agent}

  initialise converter $\mathcal{C}\gets\mathrm{BioinfoMCP}(\mathcal{M},\mathcal{K})$\;
  \eIf{$h=1$}{
    get manual info from local .pdf file: $\mathcal{D}\gets\mathrm{pdf2text}(M)$
  }{
    get manual info from the command line: $\mathcal{D}\gets\mathrm{subprocess}([T,\,\texttt{--help}])$\;
  }
  \Repeat{\normalfont syntax = valid}{
    $\mathcal{P}\gets\mathrm{generate\_prompt}(T,\mathcal{D})$\;
    $\mathcal{R}\gets\mathrm{parse\_mcpcode}(\mathcal{C}_{code})\gets(flag,err,\mathcal{C}_{code})\gets\mathrm{LLM}(\mathcal{P})$\;
    \If{$flag=0$}{
    $\mathcal{P}_{err}\gets\mathrm{generate\_prompt\_from\_error(T, \mathrm{err}, \mathcal{C}_{code})}$
    $\mathcal{R}\gets\mathrm{refine\_after\_feedback}(T,\mathcal{C}_{code},err)$\;
    }
  }
  write $O/\texttt{app}/T\texttt{\_server.py}$ with $\mathcal{C}_{code}$\;
  $\mathcal{F}_{\mathrm{docker}}\gets\mathrm{dockerfile\_content}(T)$\;
  write $O/\texttt{Dockerfile}$ with $\mathcal{F}_{\mathrm{docker}}$\;
  $\mathcal{F}_{\mathrm{compose}}\gets\mathrm{dockercompose\_content}(T)$\;
  write $O/\texttt{docker-compose.yml}$ with $\mathcal{F}_{\mathrm{compose}}$\;
  \Return{AI Agent configuration info}\;
\end{algorithm*}

The conversion process of BioinfoMCP is divided into three stages: preparation, execution, and delivery. The \textbf{preparation} stage consists of preparing the manual and the available options from the bioinformatics tool, which will be used by the LLM model backbone later in the execution stage. There are currently two available options to provide these manuals: downloading the documentation into a PDF version or accessing them using the help flag option by the tools, such as \textit{--help} or \textit{-h}.  Due to its reliance on the manual, the quality and clarity of the manual can have a substantial impact on the robustness of the generated MCP server. Moreover, BioinfoMCP-converted MCP servers are built upon the FastMCP 2.0 \cite{fastmcp} framework, which facilitated an efficient yet simple procedure to have production-level MCP servers. The base requirements for the manual are that it has a clear structure on how to execute it in the command line, together with a complete list of the flags or tags that can be utilized to modify the command execution. During the \textbf{execution} stage, BioinfoMCP will proceed with MCP server code generation with the assistance of an LLM model backbone. The system then parses and extracts code blocks from the LLM’s output. This includes detecting Python script blocks, which are then evaluated against two primary failure conditions: no code detected and syntax error. If failure did occur, it would undergo a re-generation and refinement step until it is deemed successful and proceed to the delivery stage. At the \textbf{delivery} stage, with the already refined MCP server code, BioinfoMCP Converter will pack it together with complementary files that are necessary to package them into a Docker Image that can later be an executable Docker Container. The conversion workflow framework that BioinfoMCP adopted is illustrated in \textbf{Figure~\ref{fig:workflow}} and laid out in detail in Algorithm \ref{algo:converter}, which can be found in the supplementary materials.

% \subsection{System Prompt Engineering}
The code generation in the execution stage was powered by an LLM model backbone, which is controlled by a system prompt. A system prompt is a set of instructions that rules the LLMs context and behavior, including output formats, safety guardrails, and rules that they must adhere to \cite{cheng2024prompt}. In order to perform its task correctly and produce robust MCP server code with appropriate structure, the system prompt must also be structured in a clear and complete manner. As illustrated in Figure \ref{fig:rtir} in the Appendix, the system prompt for BioinfoMCP Converter is constructed following the \textbf{role, task, requirements, and instructions}, but swapping the requirements and instructions sections to enhance the flow for the BioinfoMCP Converter to get a better comprehension. The \textbf{role} section clearly states BioinfoMCP Converter’s general description and how it should approach the tools conversion. After understanding its stance from the role section, the \textbf{task} declares the general direction that our Converter should approach. BioinfoMCP Converter then gets a step-by-step route in the \textbf{instructions} section, from which packages should be imported, parameter and file handling, subprocess execution, how to return the structured output onto the next step, and the final code format to emphasize the generated result structure. Lastly, the \textbf{requirements} are strict rules that the generated results must comply with to work in accordance with the MCP server guidelines and to guarantee robustness.

% \subsection{Configuration Setup}
%In terms of the LLM backbone, BioinfoMCP Converter can flexibly engage with models that support having a system and user prompt. For the results reported in this study, the author utilized the GPT 4.1-mini model to generate the results.

\subsection{BioinfoMCP Benchmark}
\label{sec2-benchmark}

While autonomous and flexible tool execution capabilities are necessary, they are not sufficient on their own. These tools must also be capable of executing at the appropriate contextual moments while ensuring the accuracy of their outputs. The BioinfoMCP Benchmark plays a prevalent role in ensuring the smoothness of these newly-generated MCP servers.

BioinfoMCP Benchmark follows a particular prompting structure (see supplementary), which is then sent to AI agents to call tools, generate results, and summarize key findings. Our benchmarking strategy is divided into two parts: first, every MCP server was tested independently, and second, the AI agents were tested to execute a set of bioinformatics tasks using the assistance of these MCP servers.

For the \textbf{first} part of the evaluation, MCP servers were tested to determine whether they 1) execute without encountering any non-internal tool errors, and 2) output results as expected if they were being run manually. This step is crucial to guarantee that the building blocks of future, more complicated tasks can be executed rigorously. Hence, scientists can then just focus on completing multifaceted analyses instead of validating the reliability of each step of tool-calling. In terms of the computing power utilized, for the local AI Agent, the testing was done in a remote computer with 64 GB of RAM, while for Claude Desktop and Cursor, the testing was done in a 16 GB memory local computer. After validating each tool, we then designed experiments to determine how these tools are utilized in actuality, which became the \textbf{second} part of our benchmark. The experiments were designed as shown in Table \ref{tab:experiment}, and the prompt is structured so that the AI agent can run commands for a particular task to execute an end-to-end pipeline from a genomic file. Afterwards, the AI agent is asked to summarize the results from the commands that were run. 

%\subsection{BioinfoMCP Platform}
% \label{sec2-platform}

\begin{table*}[!htb]
    \caption{BioinfoMCP Converted MCP Servers versatility assessed by the number of code lines (NCL) and robustness evaluated by using three widely-utilized AI Agents: a) Local AI Agent (LAI) b) Claude Desktop (CD) and c) Cursor (CR).}
    \label{tab:result}
    \centering
    \begin{tabular}{llllll}
        \toprule
        \multirow{2}{*}{Tool Name} & \multirow{2}{*}{NCL} & \multicolumn{3}{c}{AI Agents} \\
        \cmidrule(r){3-5} & & LAI & CD & CR\\
        \midrule
        bcftools \cite{bcftools} & 1081 & \ding{51} & \ding{51} & \ding{51}\\
        Bedtools:coverage \cite{bedtools} & 162 & \ding{51} & \ding{51}  & \ding{51}\\
        Bedtools:intersect \cite{bedtools} & 197 & \ding{51} & \ding{51} & \ding{51}\\
        Bowtie2 \cite{bowtie} & 665 & \ding{51} & \ding{51} & \ding{51} \\
        BWA \cite{bwa} & 464 & \ding{51} & \ding{51} & \ding{51} \\
        Cell-ranger \cite{cell-ranger} & 1297 & \ding{53} & \ding{53} &  \ding{53} \\
        Cutadapt \cite{cutadapt} & 400 & \ding{51} & \ding{51} &  \ding{51} \\
        DeepTools:bamCoverage \cite{deeptools, deeptools2} & 79 & \ding{51} & \ding{51} & \ding{51} \\
        DeepTools:computeGCBias \cite{deeptools, deeptools2} & 81 & \ding{51} & \ding{51} & \ding{51}\\
        DeepTools:correctGCBias \cite{deeptools, deeptools2} & 120 & \ding{51} & \ding{51} & \ding{51} \\
        DeepTools:plotCorrelation \cite{deeptools, deeptools2} & 59 & \ding{51} & \ding{51} & \ding{51}  \\
        fastp \cite{fastp} & 362 & \ding{51} & \ding{51} &  \ding{51}\\
        FastQC \cite{fastqc} & 106 & \ding{51} & \ding{51} &  \ding{51} \\
        UCSC-FaToTwoBit \cite{ucsc-fatotwobit} & 78 & \ding{51} & \ding{51} & \ding{51}\\
        Flye \cite{flye} & 151 & \ding{51} & \ding{51} & \ding{51} \\
        freebayes \cite{freebayes} & 494 & \ding{51} & \ding{51} & \ding{51} \\
        GATK:ApplyBQSR \cite{gatk} & 294 & \ding{51} & \ding{51} & \ding{51}\\
        GATK:BaseRecalibrator\cite{gatk} & 335 & \ding{51} & \ding{51} & \ding{51}\\
        GATK:HaplotypeCaller \cite{gatk} & 536 & \ding{51} & \ding{51} & \ding{51}\\
        GATK:SelectVariants  \cite{gatk} 
        & 617  & \ding{51} & \ding{51} & \ding{51}\\
        Gunzip & 424 & \ding{51} & \ding{51} & \ding{51}\\
        HISAT2 \cite{hisat2} & 523 & \ding{51} & \ding{51} & \ding{51}\\
        Kallisto  \cite{kallisto} & 575 & \ding{51} & \ding{51} & \ding{51} \\
        MACS3:callpeak \cite{macs3} & 324 & \ding{51} & \ding{51} & \ding{51}\\
        MACS3:hmmratac \cite{macs3} & 181 & \ding{51} & \ding{51} & \ding{51}\\
        Minimap2 \cite{minimap2} & 275 & \ding{51} & \ding{51} & \ding{51} \\
        MAFFT \cite{mafft} & 641 & \ding{51} & \ding{51} & \ding{51} \\
        MEME \cite{meme} & 462 & \ding{51} & \ding{51} & \ding{51} \\
        MultiQC \cite{multiqc}& 139 & \ding{51} & \ding{51} & \ding{51} \\
        Qualimap \cite{qualimap} & 613 & \ding{51} & \ding{51} & \ding{51} \\
        Quast \cite{quast} & 560 & \ding{51} & \ding{51} & \ding{51}\\
        Salmon \cite{salmon} & 331 & \ding{51} & \ding{51} & \ding{51}  \\
        SamTools \cite{samtools} & 540 & \ding{51} & \ding{51} & \ding{51} \\
        Seqtk \cite{seqtk} & 84 & \ding{51} & \ding{51} & \ding{51}  \\
        SPAdes \cite{spades} & 477 & \ding{51} & \ding{51} & \ding{51}\\
        STAR \cite{star} & 504 & \ding{51} & \ding{53} & \ding{53}\\
        Trim-galore \cite{trim-galore}& 403 & \ding{51} & \ding{51} & \ding{51} \\
        Trimmomatic \cite{trimmomatic} & 333 & \ding{51} & \ding{51} &  \ding{51}\\
        \bottomrule
    \end{tabular}
    
\end{table*}
\begin{figure*}[!htb]
    \centering
    \includegraphics[width=1\columnwidth, alt={Examples of running an MCP server, which is FastQC, using an AI Agent, here there are a User Prompt, which then intrigue an MCP server request, and then the AI Agent receives a server response, both in a form of JSON files. Afterwards, the AI Agent can then response according to the user prompt.}]{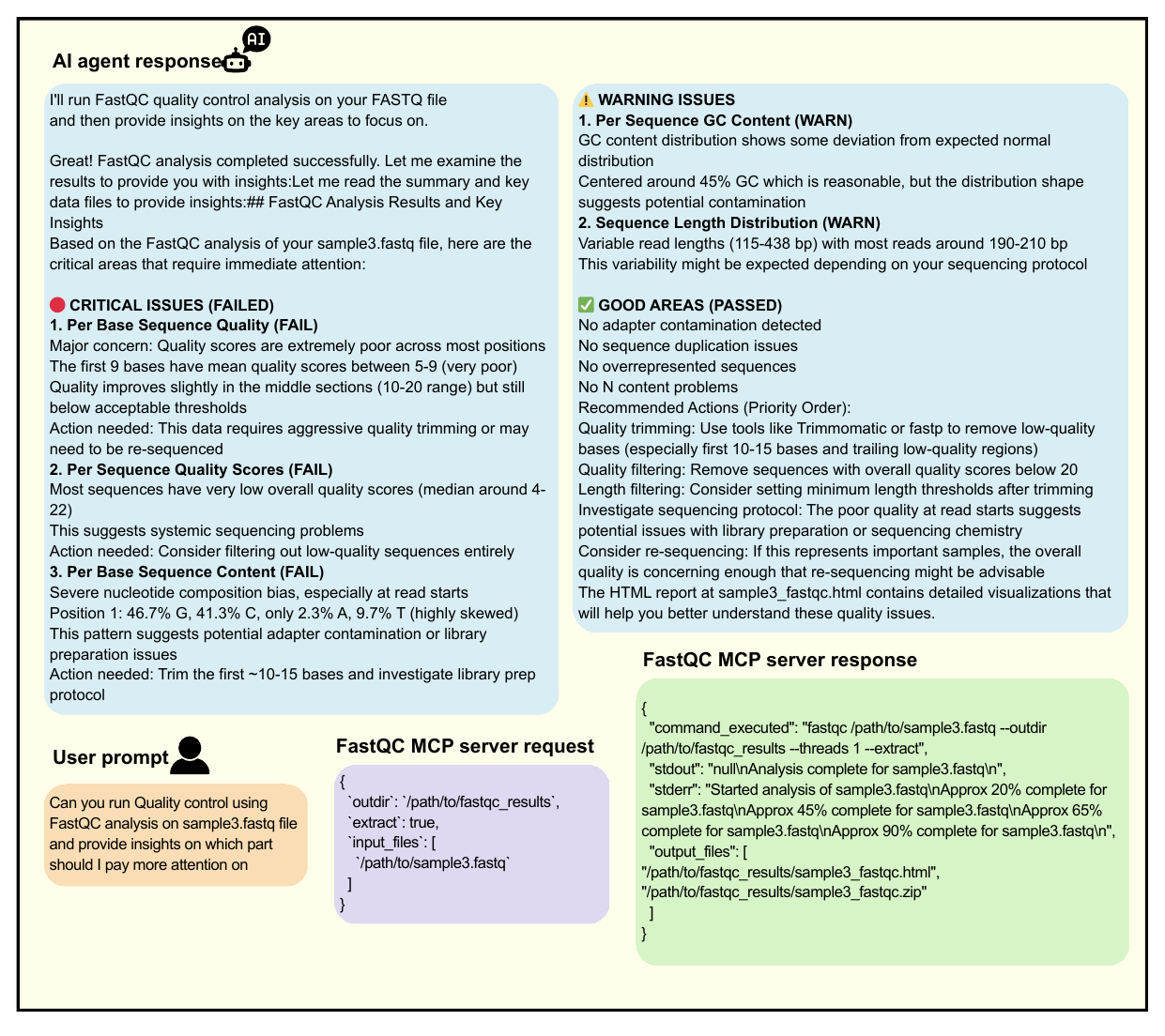}
    %\includesvg[width=1\columnwidth]{Images/exec-fastqc.svg}
    \caption{Mechanism of how AI agent (Claude Sonnet 4) make a request to an MCP Server (FastQC in this case) and obtained the response back.}
    \label{fig:example}
\end{figure*}

\begin{figure*}[!htb]
    \centering
    \includegraphics[width=1\columnwidth, alt={Running the MCP servers with a user prompt "I already execute the STAR tool and the results are in /path/to/STAR/result, now I want to move onto the next step in RNA-seq pipeline, which first is to sort and then measure the quality with respect to the annotated file", then the AI Agent execute it with samtools, qualimap, samtools, qualimap, MultiQC, and then FastQC. The AI Agent then summarize the results.}]{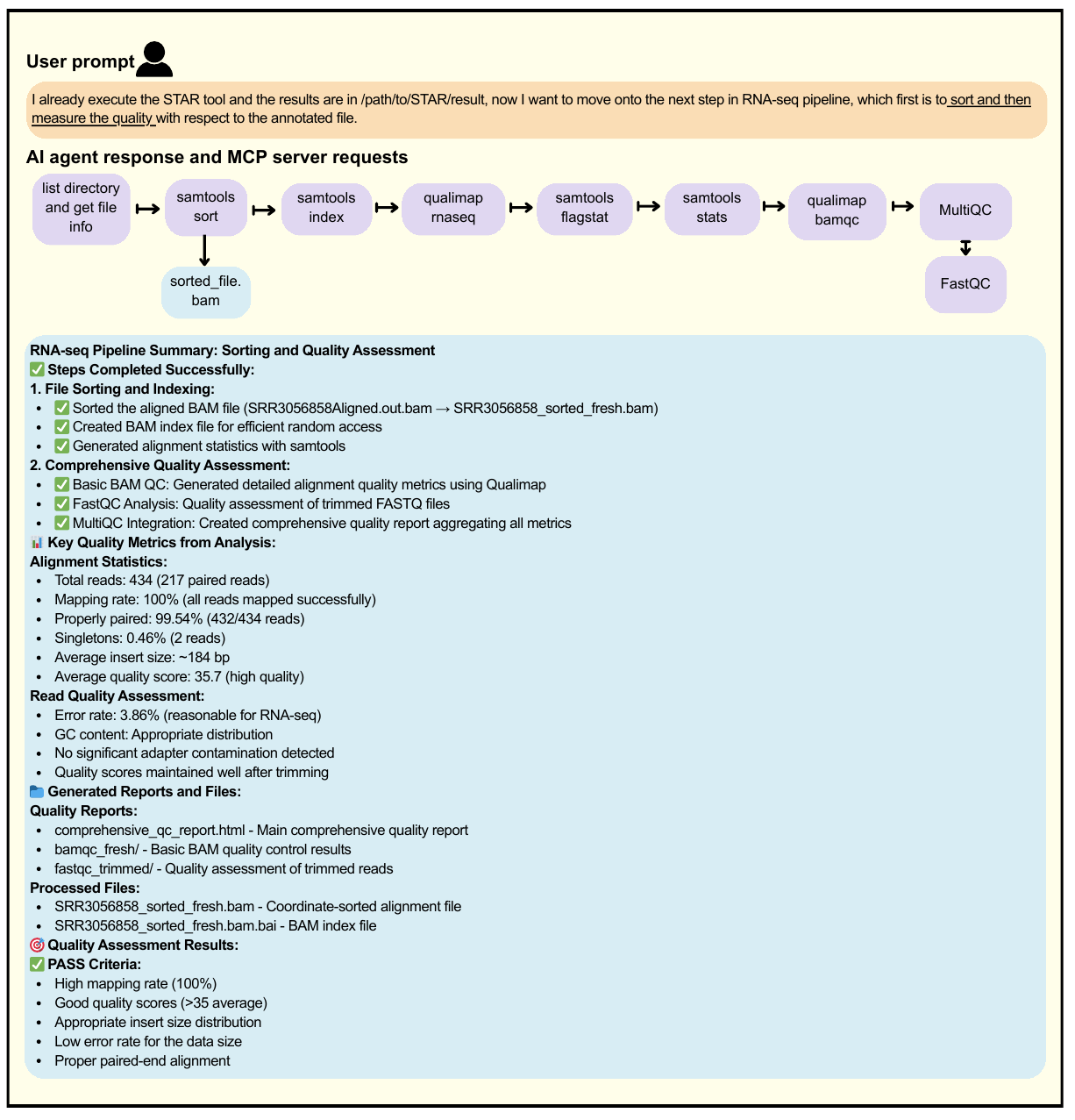}
   \caption{Finding the differentially expressed genes as part of the RNA-seq pipeline with using the MCP servers produced by BioinfoMCP Converter.}
    \label{fig:pipe1}
\end{figure*}

\section{Results}

% Present your findings.
% Use figures, tables, and stats. Explain what they mean briefly.

\subsection{Converting Bioinformatics Tools into MCP Servers with BioinfoMCP Converter}
Just by providing an official manual provided by the tool, BioinfoMCP Converter can extract essential information from the manual passed onto the platform and directly transform it into an MCP server. It also adopts its standardized structure that was depicted by the official MCP documentation, which consists of tools, resources, and prompts, so it is able to connect with any AI Agents that have an MCP client attached to it. Moreover, BioinfoMCP Converter generates the MCP servers with a detailed description, capturing the utility of each one of the tags or flags. Thus, making it possible for any combination of attributes to be run on the commands. 

Although the converter-powering generative model is alterable, this study also has tested on several models to compare the conversion efficacy of the FastQC tool as a baseline for future model selection. As shown in Table \ref{tab:compare}, GPT-4.1-mini completed the conversion task in just 13.7 seconds, making it the second-fastest overall. From the perspective of cost-efficiency, its total cost is only marginally higher than the cheapest option (Deepseek-chat \cite{deepseek}) but remains an order of magnitude less expensive than GPT-4o \cite{gpt4o}. Furthermore, GPT-4.1-mini \cite{gpt41} delivered an 88-line implementation, which strikes the ideal balance between the verbose 194-line output of Gemini-2.5-flash \cite{gemini} and the overly concise 48-line version from GPT-4o-mini. Some factors that might contribute to this difference for instance is the context window, on which GPT-4o-mini  has of 128 thousand, while GPT-4.1-mini  has around one million. This comparison concludes that the model backbone also played an important role in determining the comprehensiveness and quality of the generated MCP servers. 

\begin{table}[!htb]
    \caption{Comparison between the performance of BioinfoMCP Converter in Converting FastQC tool onto an MCP server using different Backends in terms of the conversion time (in seconds), number of code lines (NCL), number of tokens (NT), and the conversion cost (in dollars)}
    \label{tab:compare}
    \centering
    \begin{tabular}{lllll}
    \toprule
        Converter Backend & Time (s) & NCL & NT & Cost (\$) \\
    \midrule
        GPT-4.1-mini \cite{gpt41} & 13.7188 & 88 & 879 & 0.01222 \\
        GPT-4o-mini \cite{gpt4o}& 12.1285 & 48 & 484 & 0.01151 \\
        GPT-4o \cite{gpt4o}& 13.7019 & 72 & 801 & 0.11272 \\
        Gemini-2.5-flash \cite{gemini} & 27.4415 & 194 & 3112 & 0.02834 \\
        Deepseek-chat \cite{deepseek} & 52.9808 & 106 & 1085 & 0.00996 \\
    \bottomrule
        
    \end{tabular}
    
\end{table}

The extended context window of GPT-4.1 mini proved particularly advantageous for tools with extensive documentation, such as SPAdes \cite{spades} and Samtools \cite{samtools}, where comprehensive parameter sets and complex usage patterns required substantial contextual understanding. In contrast, GPT-4o mini, while more computationally efficient, occasionally truncated or simplified complex parameter descriptions when processing lengthy manuals. This trade-off between computational cost and conversion quality suggests that model selection should be tailored to the complexity of the target bioinformatics tool and the desired level of detail in the resulting MCP server.

In terms of the converted results, BioinfoMCP Converter has successfully converted 38 different tools, which have various options (details regarding each tool can be seen in Table \ref{tab:result}). In addition, for versatile multi-function tools, such as GATK \cite{gatk} or Deeptools \cite{deeptools, deeptools2}, BioinfoMCP Converter operates in each of the subtools to guarantee that every functionality is captured inside the MCP server, although with the extensive context window, it is probably not sufficient to apprehend the
intricate details. The transformation process from raw command-line documentation to fully functional MCP servers demonstrated remarkable efficiency, averaging 40 seconds per tool and requiring no more than two minutes even for complex applications such as bcftools and cell-ranger. In terms of prospective tools, BioinfoMCP Converter, as it works automatically, will always be ready to convert those newly-released tools in the future.

\subsection{Evaluating AI Agent in Agentic Bioinformatics with BioinfoMCP Benchmark}

\subsubsection{Individual MCP servers benchmarking}

Individual MCP servers are tested in different scenarios, from local AI agent, Claude Desktop \cite{claude} and Cursor \cite{cursor}, and observed whether these agents can perform their intended utility or not. However, some tools in the bioinformatics domain are memory or time-consuming, which cause unintended failures that were not directly caused by the MCP servers or the tool-calling procedure itself, as AI agents were able to connect to the intended tool with precise commands. The efficacy with respect to the individual tool testing, including the number of code lines to demonstrate completeness, is depicted in Table \ref{tab:result}.

\subsubsection{Pipeline Execution benchmarking}
\begin{table*}[!htb]
    \centering
    \caption{Experiment Design of BioinfoMCP Benchmark for MCP server and server-agent interaction testing.}
    \small
    
    \begin{tabular}{p{2.5cm} p{4cm} p{1.84cm} p{6.5cm} p{1.7cm}}
        \toprule
        Pipeline Analysis Name & Task Name & Result Diagram & Bioinformatics Tools Utilized & Time Required \\
        \midrule
        RNA-seq & Find differentially expressed genes  & \ref{fig:pipe1} & FastQC, samtools, Qualimap, MultiQC & 4\\
        WGS & Genome assembly & \ref{fig:pipe2} & FastQC, fastp, SPAdes, Quast, MultiQC & 5\\
        ChIP-seq & Motif discovery for binding sites & \ref{fig:pipe3} & FastQC, Bowtie2, samtools, MACS3, Deeptools, MultiQC, R (GenomicRanges, GenomicAlignment, Rsamtools) & 11\\
        ATAC-seq & Identifying open chromatin region & \ref{fig:pipe4} & FastQC, Trim-galore, Bowtie2, samtools, MACS3, MultiQC & 7\\
        WGS/WES & Somatic SNV calling & \ref{fig:pipe5} & FastQC, fastp, Bowtie2, samtools, GATK, Freebayes, bcftools & 9\\
        \bottomrule
    \end{tabular}
    \label{tab:experiment}
\end{table*}

After connecting to an AI agent, each MCP server can interact with the others to perform a complex task that requires multiple execution steps, which are the baseline of bioinformatics-pipelined tasks. For instance, to find the differentially expressed genes in a RNA-seq pipeline, it will first conduct a pre-alignment quality control and pre-processing over the raw fastq files, then moving on to alignment to reference genome indexes, followed by post-alignment quality assessment, read quantification at gene or transcript level, normalization of count data, and finally statistical analysis to identify genes with significant expression differences between experimental conditions \cite{rnaseq-survey}. With the assistance of MCP servers, AI agents can call subsequent tools according to the user's requirements. As shown in Figure \ref{fig:pipe1}, BioinfoMCP-generated MCP servers, together with the help of the filesystem MCP server, AI agents are able to carry out instructions according to the user prompt in a reliable manner. Suppose an error occurs during the operations (such as the Qualimap:rnaseq execution shown in Figure \ref{fig:pipe1}). In that case, these AI agents can adapt accordingly by understanding the error message and then make consequent modifications before stepping forward to the next call. Other experiments can be seen in the supplementary materials of this study.

%Pipeline Name, Task Name, Which tool it executes with and whether it is an appropriate tool for the corresponding task,  Whether or not it can generate appropriate commands, whether it can execute the commands, whether it gives appropriate explanation and next steps, time required

\subsection{MCP servers returns a comprehensible request and result connection for AI agents and human to interpret}

BioinfoMCP Converter-generated MCP servers are not constructed just as a jaggy bridge between these useful-but-rigid tools to connect to cutting-edge AI agents, but they are also created to lay a smooth foundation for AI agents to easily call these tools, and in turn, enable scientists to get the results as intended easily. In terms of sending the MCP requests, these clients are able to easily call the MCP servers as these servers are provided with a complete set of parameter options to call, on which, if it is an optional one, an appropriate default value will be given accordingly. After executing the request and attaining the result, these MCP servers will deliver a constructed output of three essential information: the \textit{command that was run}, the \textit{stderr}, and the \textit{stdout}. As depicted in Figure~\ref{fig:example}. This constructed output format assists AI agents in not only executing the tools effortlessly but also collecting insights from the result of the executed command and interpreting it for the user's understanding. Without sending out this constructed output, AI agents can only run without knowing their execution status or outcome.

With the functionality that BioinfoMCP Converter-generated servers have provided, it will not only utilize the bioinformatics tools to its full potential by bridging the gap for AI agents to use these tools, but it will also boost productivity for experts, as they do not have to run abstract command-line interface (CLI) commands for tools execution anymore and can execute tools by utilizing human-machine interaction. Moreover, by combining the strength of AI agents' thinking capability, bioinformaticians will have another set of interpretations and be able to get a glimpse of the result before diving deep into the report.

\section{Discussion and Future Works}

The BioinfoMCP platform has paved the way for the conversion of any bioinformatics tool into a robust executable MCP server by utilizing the tool's documentation without any requirement for human intervention. With an MCP server readily available, these tools can be executed on AI agents that have MCP clients attached to them, so users only need to instruct these AI agents with human language instead of abstract CLI commands. With a rigorous system prompt, BioinfoMCP Converter-generated MCP servers have a complete set of comprehensible parameters that can then be translated to CLI commands so AI agents can send a request without understanding the underlying detailed structure of each tool. This study has also proven that the BioinfoMCP-Converted MCP server's reliability in executing bioinformatic tools operations. This advancement has also made it possible for AI agents to send multiple requests sequentially according to the user request, which will be significantly beneficial in bioinformatics domain tasks.

However, the present version of BioinfoMCP Converter necessitates manual retrieval and interpretation of help documentation for integrated third-party tools, commonly accessed via command-line flags such as –help or manually retrieving their tool documentation online. To streamline this process, we plan to design an automated system where users need only specify the tool name; this framework will autonomously fetch the relevant documentation by programmatically invoking the tool's built-in help function (e.g., tool --help) or, if necessary, retrieving it from curated online sources, which eliminates manual intervention while ensuring accurate, up-to-date usage guidelines are available for downstream operations. It is also important to note that in terms of the AI agent capabilities, there is still a toil in conducting end-to-end task completion since AI agents are still unable to handle tools that require extensive memory or runtime such as STAR as current AI agents are still unable to be connected to Graphic Processing Units (GPUs), which make these heavy workloads still require manual human-assisted execution. Hence, while the MCP servers demonstrate functional robustness across different AI agent integrations, deployment considerations such as memory allocation and execution timeouts must be adequately configured to ensure reliable performance in real-world applications.

%Additionally, for tools that inherently were in a form of R package or Python packages, such as Seurat or DeSEQ2 for the former and PyDeSEQ2 (DeSEQDataset) or scanpy for the latter, it is still necessary to run through an R or Python integrator, such as ClaudeR \cite{clauder}, as current 

Nonetheless, the invention of BioinfoMCP Converter and its complementary Benchmark has addressed the gap between the incompatibility of bioinformatics tools to be executed on recent AI agents. At the same time, the automated approach has eradicated the manual conversion bottleneck and has made this technique reliable for future usability of future tools. The BioinfoMCP family will make a significant contribution to bioinformaticians' productivity improvement. With further support for BioinfoMCP development, the utilization of AI technology in scientific experiments will be revolutionized.

%\section{Code Availability}
%The BioinfoMCP platform is available at ... and the corresponding codes are available publicly at ...
\section{Acknowledgments}
The authors from this research work were supported by grants from the 28th Undergraduate Research Award from the Chinese University of Hong Kong, Shenzhen (CUHK Shenzhen) and also from award number UDF01004172.

\section{Authors Contribution}
F.W. and Z.C. contributed equally to this work. J.Z. with F.W. conceptualized the platform. F.W. developed the code. Z.C. developed and tested the local AI Agent. F.W. and Z.C. performed experiments. F.W. and J.Z. wrote the manuscript. F.W., Z.C., and J.Z. revised critical parts of the manuscript. J.Z. performed supervision. F.W. and J.Z. responsible for funding acquisition.

\section{Code availability}
The BioinfoMCP platform is publicly available at https://github.com/florensiawidjaja/BioinfoMCP

{
\bibliographystyle{IEEEtran}
\bibliography{main}
}
\newpage
\section{Supplementary Materials}

\subsection{More Function Calling Examples}
\subsubsection{Bowtie2 Align}
\textbf{MCP Server Request:}
\begin{lstlisting}[style=jsonStyle]
{
  `local`: true,
  `threads`: 4,
  `end_to_end`: false,
  `index_base`: `/data/bowtie2_GRCh38_index`,
  `sam_output`: `/data/SRR8405197_aligned_v2.sam`,
  `mate1_files`: `/data/trimmed/SRR8405197_1_val_1.fq.gz`,
  `mate2_files`: `/data/trimmed/SRR8405197_2_val_2.fq.gz`,
  `very_sensitive_local`: true
}
\end{lstlisting}
\textbf{MCP Server Response:}
\begin{lstlisting}[style=jsonStyle]
{
  "command_executed": "bowtie2 -x /data/bowtie2_GRCh38_index -1 /data/trimmed/SRR8405197_1_val_1.fq.gz -2 /data/trimmed/SRR8405197_2_val_2.fq.gz -S /data/SRR8405197_aligned_v2.sam -q --very-sensitive-local -N 0 --dpad 15 --gbar 4 --local --ma 2 --mp 6,2 --np 1 --rdg 5,3 --rfg 5,3 -D 15 -R 2 -I 0 -X 500 --fr --met 1 -p 4 --seed 0",
  "stdout": "",
  "stderr": "4841 reads; 
    of these:  4841 (100.00%) were paired; 
    of these: 
    3655 (75.50%) aligned concordantly 0 times
    394 (8.14%) aligned concordantly exactly 1 time
    792 (16.36%) aligned concordantly >1 times
    ---- 
    3655 pairs aligned concordantly 0 times; 
    of these:      
    10 (0.27%) aligned discordantly 1 time
    ----
    3645 pairs aligned 0 times concordantly or discordantly;
    of these:
    7290 mates make up the pairs;
    of these:
    6941 (95.21%) aligned 0 times
    201 (2.76%) aligned exactly 1 time
    148 (2.03%) aligned >1 times
    28.31% overall alignment rate",
  "output_files": [
    "/data/SRR8405197_aligned_v2.sam"
  ]
}


\end{lstlisting}
\vspace{10em}

\subsubsection{Cutadapt}

\begin{samepage}
\textbf{MCP Server Request:}
\begin{lstlisting}[style=jsonStyle]
{
  "cores": 4,
  "adapter": "AGATCGGAAGAGCACACGTCTGAACTCCAGTCA",
  "input_file": "/data/SRR8405197.fastq",
  "output_file": "/data/SRR8405197_trimmed.fastq",
  "minimum_length": "20",
  "quality_cutoff": "20"
}
\end{lstlisting}
\end{samepage}
\textbf{MCP Server Response:}
\begin{samepage}
\begin{lstlisting}[style=jsonStyle]
{
  "command_executed": "cutadapt -j 4 -a AGATCGGAAGAGC... [truncated]",
  "stdout": "This is cutadapt 5.1 with Python 3.10.0\n...",
  "stderr": "",
  "output_files": ["/data/SRR8405197_trimmed.fastq"]
}
\end{lstlisting}
\end{samepage}

\subsubsection{Samtools Sort}
\begin{samepage}
\textbf{MCP Server Request:}

\begin{lstlisting}[style=jsonStyle]
{
  `threads`: 4,
  `input_file`: `/data/SRR8405197_aligned_v2.bam`,
  `output_file`: `/data/SRR8405197_sorted_v2.bam`
}
\end{lstlisting}
\end{samepage}
\begin{samepage}
\textbf{MCP Server Response:}
\begin{lstlisting}[style=jsonStyle]
{
  "command_executed": "samtools sort -o /data/SRR8405197_sorted_v2.bam -O bam -@ 4 /data/SRR8405197_aligned_v2.bam",
  "stdout": "",
  "stderr": "[bam_sort_core] merging from 0 files and 4 in-memory blocks...\n",
  "output_files": [
    "/data/SRR8405197_sorted_v2.bam"
  ]
}
\end{lstlisting}
\end{samepage}
\newpage
\subsubsection{Deeptools:bamCoverage}

\begin{samepage}
\textbf{MCP Server Request:}
\begin{lstlisting}[style=jsonStyle]
{
  `bam`: `/data/analysis_output/aligned_sorted.bam`,
  `binSize`: 50,
  `outFileName`: `/data/analysis_output/coverage.bw`,
  `normalizeUsing`: `RPKM`,
  `numberOfProcessors`: 4
}
\end{lstlisting}
\end{samepage}
\begin{samepage}
\textbf{MCP Server Response:}
\begin{lstlisting}[style=jsonStyle]
{
  "command_executed": "bamCoverage -b /data/analysis_output/aligned_sorted.bam -o /data/analysis_output/coverage.bw --binSize 50 --minMappingQuality 0 --numberOfProcessors 4 --normalizeUsing RPKM",
  "stdout": "",
  "stderr": "normalization: RPKM
  bamFilesList: ['data/analysis_output/aligned_sorted.bam'] 
  binLength: 50\nnumberOfSamples: None
  blackListFileName: None
  skipZeroOverZero: False
  bed_and_bin: False
  genomeChunkSize: None
  defaultFragmentLength: read length
  numberOfProcessors: 4
  verbose: False
  region: None
  bedFile: None
  minMappingQuality: 0
  ignoreDuplicates: False
  chrsToSkip: []
  stepSize: 50
  center_read: False
  samFlag_include: None
  samFlag_exclude: None
  minFragmentLength: 0
  maxFragmentLength: 0
  zerosToNans: False
  smoothLength: None
  save_data: False
  out_file_for_raw_data: None
  maxPairedFragmentLength: 1000",
  "output_files": [
    "/data/analysis_output/coverage.bw"
  ]
}
\end{lstlisting}
\end{samepage}
\vspace{50em}
\subsubsection{SPAdes (if not successful)}

\begin{samepage}
\textbf{MCP Server Request:}
\begin{lstlisting}[style=jsonStyle]
{
  `memory`: 8,
  `careful`: true,
  `dataset`: `/data/dataset.yaml`,
  `isolate`: true,
  `threads`: 2,
  `output_dir`: `/data/spades_assembly`
}
\end{lstlisting}
\end{samepage}
\begin{samepage}
\textbf{MCP Server Response:}
\begin{lstlisting}[style=jsonStyle]
{
  "command_executed": "spades.py --isolate --careful --dataset /data/dataset.yaml -o /data/spades_assembly -t 2 -m 8",
  "stdout": "== Error ==  you cannot specify --mismatch-correction or --careful in isolate mode!
  In case you have troubles running SPAdes, you can report an issue on our GitHub repository github.com/ablab/spades
  Please provide us with params.txt and spades.log files from the output directory.",
  "stderr": "SPAdes failed with return code 255",
  "output_files": []
}
\end{lstlisting}
\end{samepage}

\subsection{AI Agent Settings (JSON format)}
\subsubsection{Connect to Python Environment}
\begin{lstlisting}[style=jsonStyle2]
{
  "command": "bash",
  "args": [
    "-c",
    "source /Users/<name>/miniforge3/etc/profile.d/conda.sh && conda activate <env-name> && python /path/to/mcp_tool/app/tool_server.py"
  ]
}
\end{lstlisting}

\subsubsection{Connect to Docker Connection}
\begin{lstlisting}[style=jsonStyle2]
{
    "command": "docker",
    "args": [
        "run", "-i", "--rm",
        "-v", "/path/to/tool/data:/app/workspace",
        "tool-server:latest"
    ]
}
\end{lstlisting}

\subsection{BioinfoMCP Benchmark Prompts}
\subsubsection{Individual Tool Execution}
\begin{lstlisting}[style=jsonStyle2]
Can you run the <tool-name> on <genomic-file-path>. Please state what commands you run and what are the outputs or results from that command.
\end{lstlisting}

\subsubsection{Pipeline Task Execution}
\begin{lstlisting}[style=jsonStyle2]
Can you run the <pipeline-name> for <task-name> to the <genomic-file-path> (and <genomic-file-path-2>). 

(If part of the pipeline has already been done, can be stated here as well with "I already run the <ran-task-name> and and the results are <results-path>")

Give appropriate explanations and summarize results at the end with a simple report.
\end{lstlisting}

\newpage

\begin{figure*}[!htb]
    \centering
    \includegraphics[width=1\columnwidth, alt={Give a detail explanation for each part of the system prompt structure for BioinfoMCP Converter, from the Role, Task, Requirements, and Instructions.}]{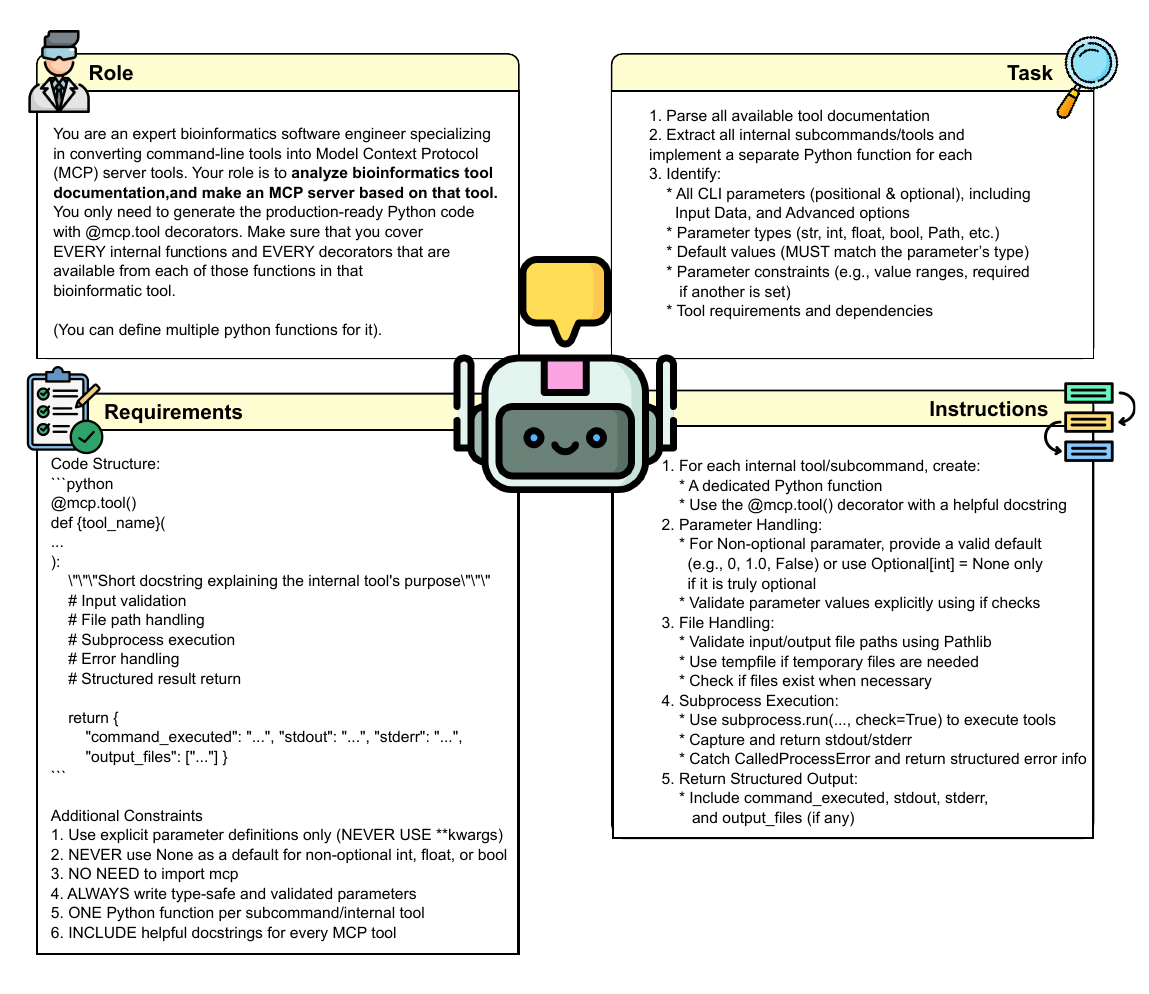}
    \caption{The system prompt structure for BioinfoMCP Converter.}
    \label{fig:rtir}
\end{figure*}

\begin{figure*}[!htb]
    \centering
    \includegraphics[width=1\columnwidth, alt={Running the MCP servers with a user prompt "Can you run the Whole Genome Sequence data analysis with the Genome Assembly task to the /path/to/SRR8405197.fastq. Give appropriate explanations and summarize results at the end with a simple report", then the AI Agent execute it with a few MCP Servers. Afterwards, the AI Agent then summarize the results.}]{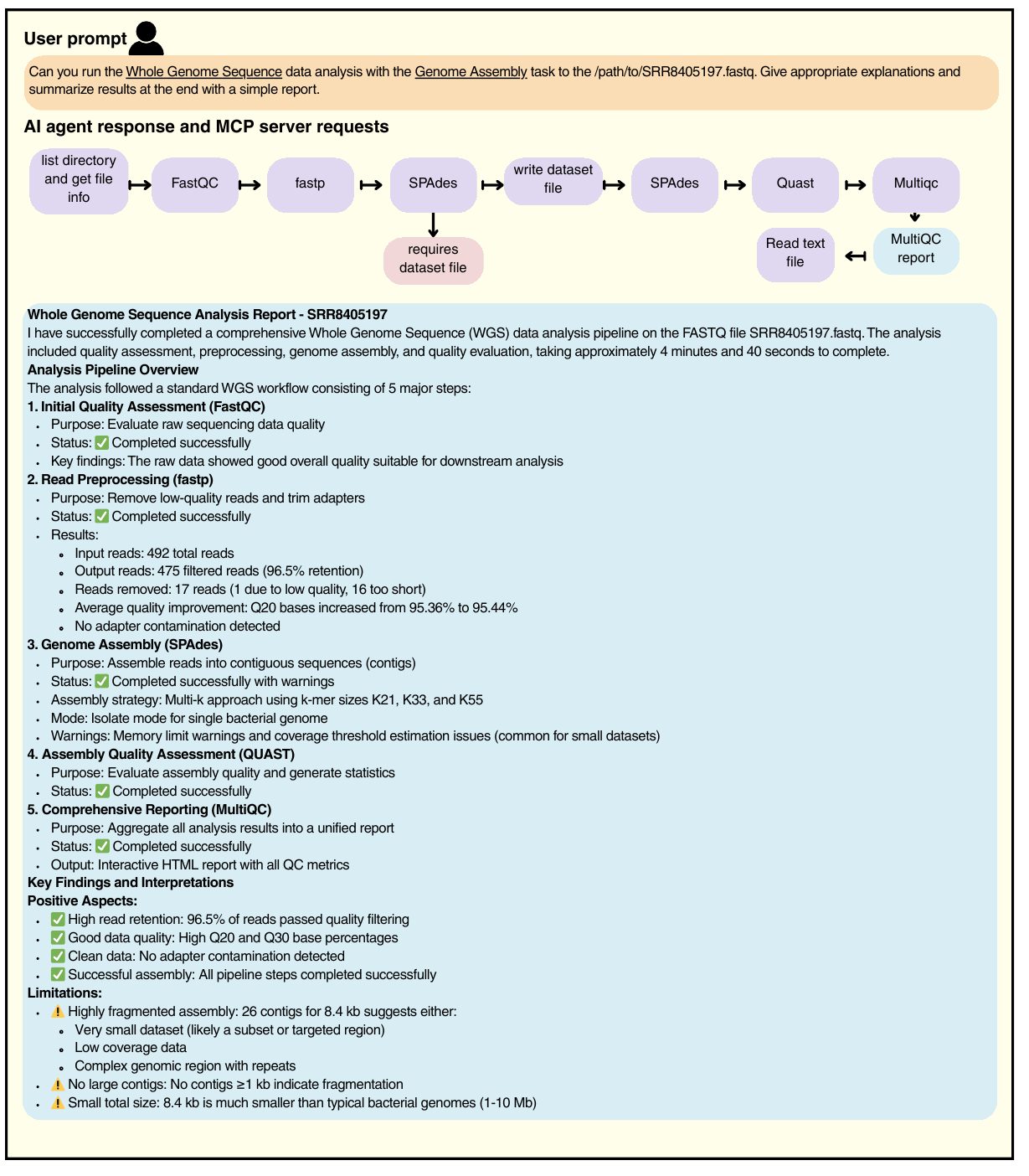}
   \caption{Running the Genome Assembly task as part of the WGS pipeline with using the MCP servers produced by BioinfoMCP Converter.}
    \label{fig:pipe2}
\end{figure*}

\begin{figure*}[!htb]
    \centering
    \includegraphics[width=1\columnwidth, alt={Running the MCP servers with a user prompt "Can you run the ChIP-seq analysis for Motif discovery for binding sites task to the /path/to/SRR8405197_1.fastq and /path/to/SRR8405197_2.fastq. Give appropriate explanations and summarize results at the end with a simple report", then the AI Agent execute it with a few MCP Servers. Afterwards, the AI Agent then summarize the results.}]{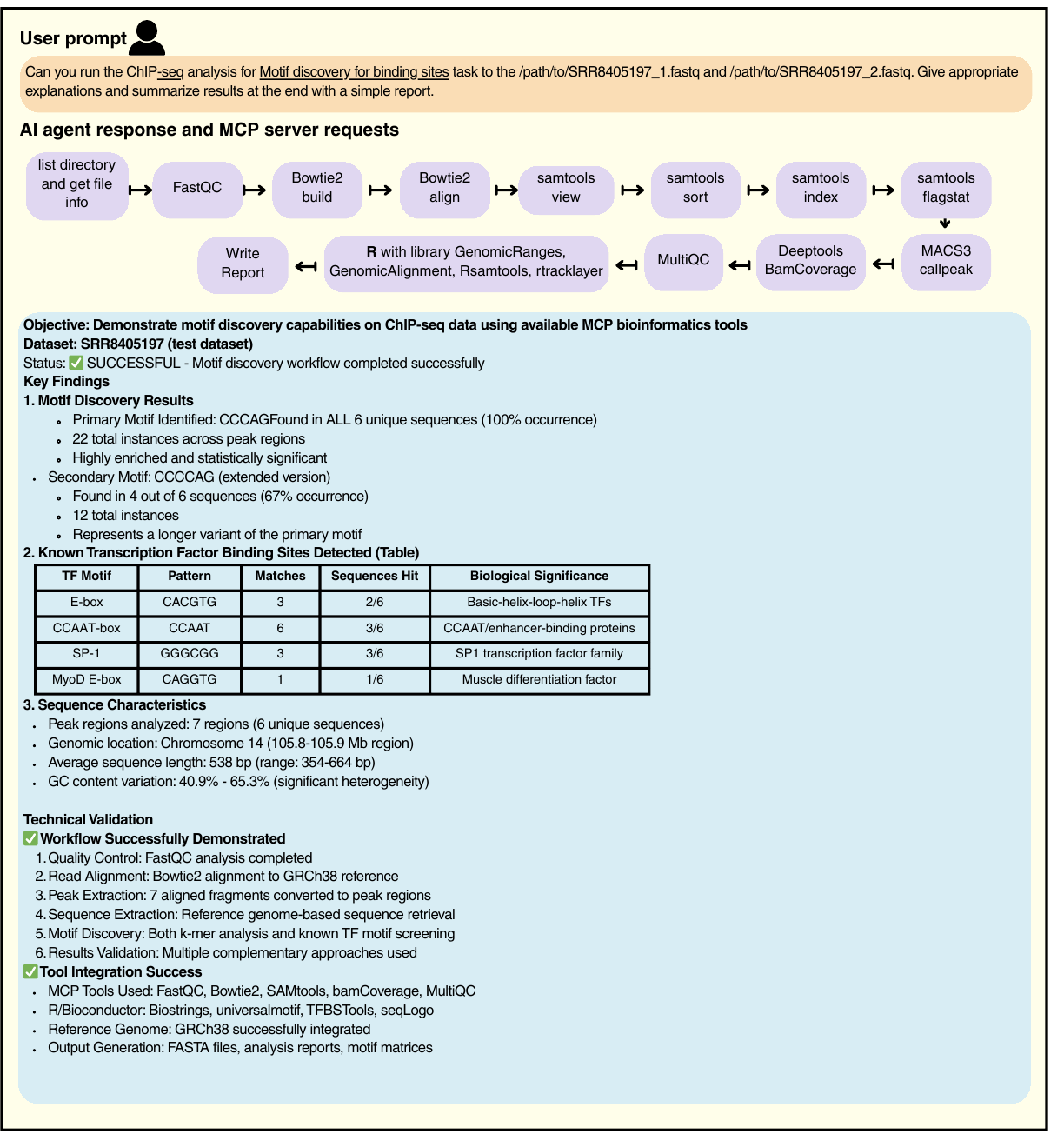}
    %\includesvg[width=\textwidth]{Images/wgsseq.svg}
    \caption{Running the Motif discovery for binding sites task as part of the ChIP-seq pipeline using the MCP servers produced by BioinfoMCP Converter.}
    \label{fig:pipe3}
\end{figure*}

\begin{figure*}[!htb]
    \centering
    \includegraphics[width=1\columnwidth, alt={Running the MCP servers with a user prompt "Can you run the ATAC-seq analysis for identifying open chromatin region task to the /path/to/SRR8405197_1.fastq and /path/to/SRR8405197_2.fastq. Give appropriate explanations and summarize results at the end with a simple report", then the AI Agent execute it with a few MCP Servers. Afterwards, the AI Agent then summarize the results.}]{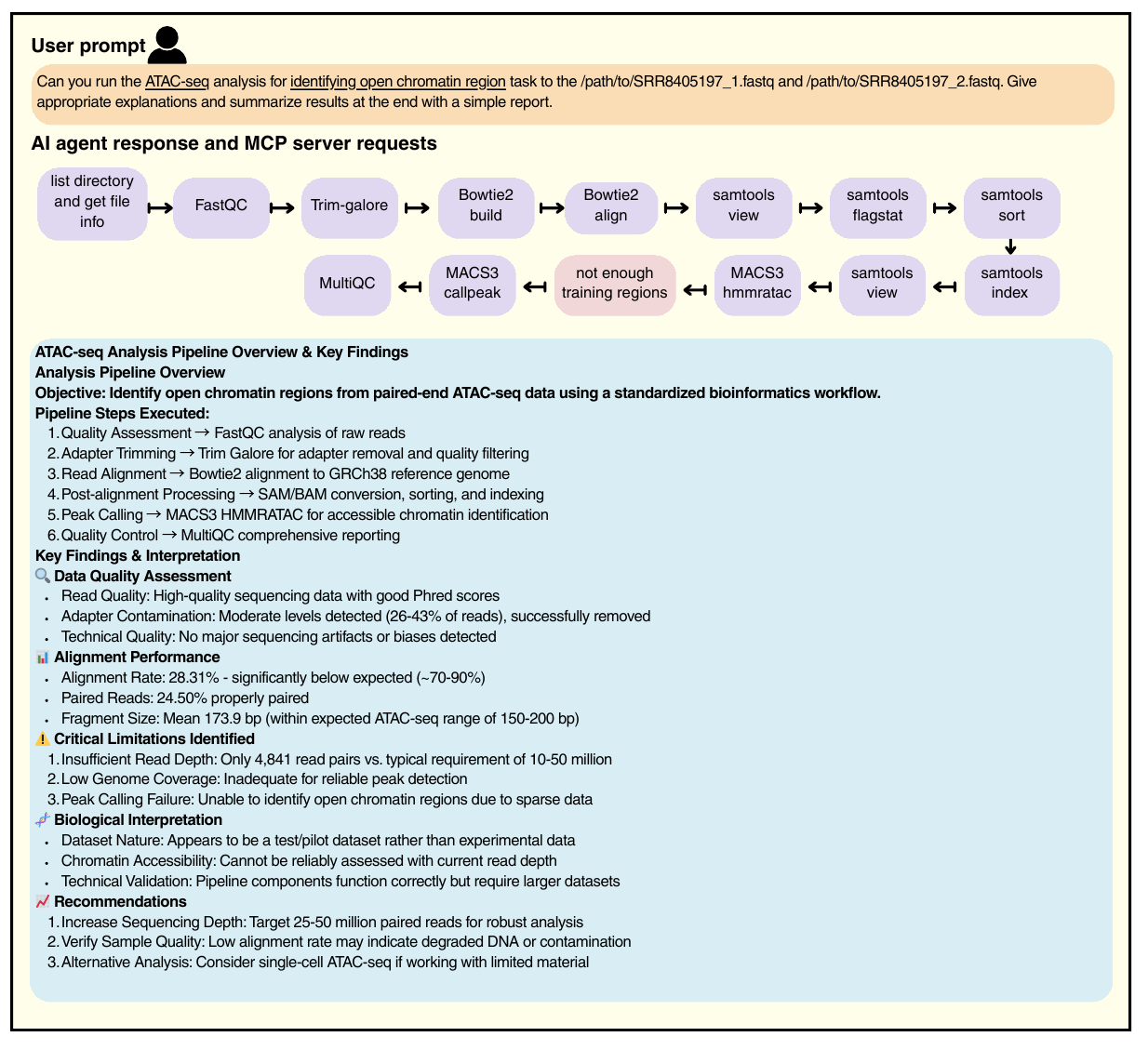}
    %\includesvg[width=\textwidth]{Images/wgsseq.svg}
    \caption{Running the identifying open chromatin region task as part of the ATAC-seq pipeline using the MCP servers produced by BioinfoMCP Converter.}
    \label{fig:pipe4}
\end{figure*}

\begin{figure*}[!htb]
    \centering
\includegraphics[width=1\columnwidth, alt={Running the MCP servers with a user prompt "Run the WGS/WES data analysis for the Somatic SNV+index task to the /path/to/SRR8405197_1.fastq and /path/to/SRR8405197_2.fastq. I already have the Reference index at /path/to/ref-index. Give appropriate explanations and summarize results at the end with a simple report.", then the AI Agent execute it with a few MCP Servers. Afterwards, the AI Agent then summarize the results.}]{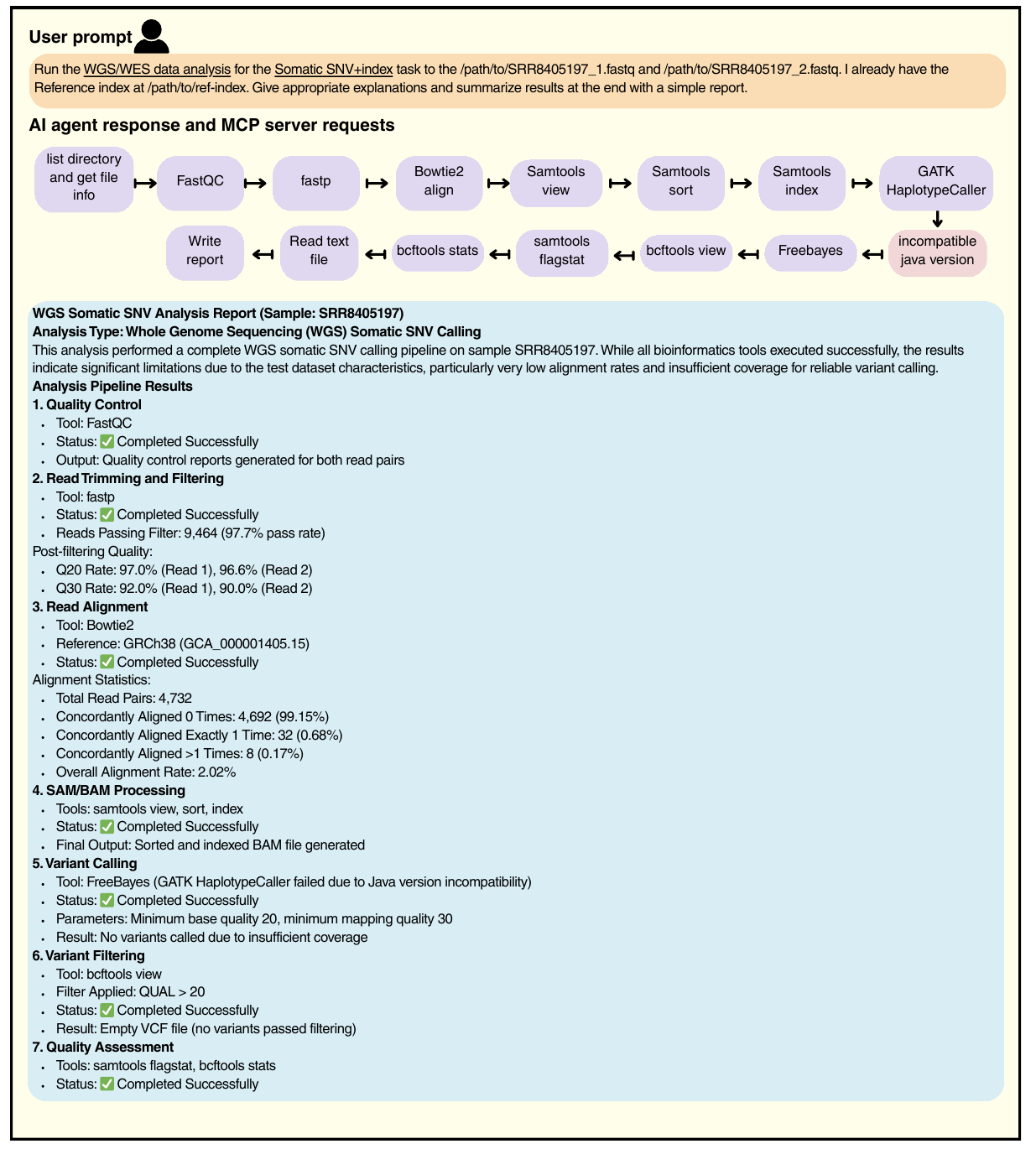}    %\includesvg[width=\textwidth]{Images/wgsseq.svg}
    \caption{Running the somatic SNV calling task as part of the WGS/WES pipeline using the MCP servers produced by BioinfoMCP Converter.}
    \label{fig:pipe5}
\end{figure*}

\end{document}